\documentclass[aps,prl,twocolumn,amsfonts,showpacs,floatfix]{revtex4}
\usepackage{epsfig}
\usepackage{times}
\usepackage{bbm}

\newcommand{\id}{{\sf 1 \hspace{-0.3ex} \rule{0.1ex}{1.52ex} \rule[-.02ex]{0.3ex}{0.1ex} }}

\begin{document}

\title{Introduction to the basics of entanglement theory in 
continuous-variable systems}

\author{J. Eisert$^{1,2}$ and M.B.\ Plenio$^2$}
\affiliation{1 Institut f{\"u}r Physik, Universit{\"a}t Potsdam, 
Am Neuen Palais 10, D-14469 Potsdam, Germany \\
2 Quantum Optics and Laser Science, Blackett Laboratory, Imperial College London,
London SW7 2BW, UK}

\begin{abstract}
We outline the basic questions that are being studied in the
theory of entanglement. Following a brief review of some of the
main achievements of entanglement theory for finite-dimensional
quantum systems such as qubits, we will consider entanglement in
infinite-dimensional systems. Asking for a theory of entanglement
in such systems under experimentally feasible operations leads to
the development of the theory of entanglement of Gaussian states.
Results of this theory are presented and the tools that have been
developed for it are applied to a number of problems.
\end{abstract}

\maketitle

\section{Introduction}

When John Bell, then a theoretical physicist at CERN, published
his now famous theorem in 1964, it was at first hardly noticed by
the scientific community. Only within the subsequent decade the
fundamental character of its content was truly appreciated
\cite{Ballentine 87}: in a sense, it fleshed out the notion of
entangled quantum systems being potentially correlated in a way
that classical systems cannot. Entanglement can indeed be viewed
as the furthest and most radical departure of quantum mechanics
from classical physics. In recent years, yet another shift in the
way quantum correlations are perceived has taken place:
entanglement has been recognized as a resource that may be
exploited to perform completely novel information processing tasks
or enhance the efficiency of known tasks. As for any resource, a
detailed understanding of its properties and the resulting
possibilities and limitations for its manipulation are an
important pre-condition for the full exploitation of its
potential. Quite naturally, the systematic investigation of
quantum entanglement in particular is a major goal of theoretical
research in quantum information science \cite{Horodecki 01,Plenio
V 98,Schumacher 00,Plenio V 01}.

This program of research has been initiated with the investigation
of the entanglement properties of bi-partite quantum systems
prepared in pure quantum states \cite{Bennett BPW 96,Lo P 01}.
Today, most characteristics of pure state entanglement are in fact
well understood \cite{Nielsen 99,Vidal 99,Jonathan P 99a,Jonathan
P 99b}. However, our understanding of mixed state entanglement is
much less complete. Most proven results are restricted to
situations where the constituent parts are qubits, quantum
two-level systems, which is due to a whole range of special
mathematical properties only satisfied by qubit systems
\cite{Vollbrecht W 99}. In contrast, the study of bi-partite
entanglement between higher-dimensional systems is much more
involved. Indeed, new types of entanglement emerge such as bound
entanglement \cite{Horodecki 97}, and one encounters a rich and
complex proliferation of types of entanglement with growing
dimension of its constituent parts.

Considering the complex structure that is emerging even in the
case of low-dimensional quantum systems, one may be tempted to
come to the conclusion that to say anything meaningful about the
entanglement of infinite-dimensional quantum systems such as field
modes of light or harmonic oscillators is an enterprise doomed to
failure. This is, however, not so: it has become clear in recent
years that generic statements on the entanglement of such quantum
systems can indeed be made. Most importantly, for a large class of
states many exact results on entanglement theory can be found,
even in cases where there is no finite-dimensional counterpart. A
brief description of the development of this theory will be the
main subject of this article. It should be noted that in this
article, a certain emphasis is put on the results that the
research groups at Imperial College and in Potsdam have
contributed to this area \cite{Remark}.

This article will be structured in the following way. After a
short introduction to some key notions of entanglement that are
valid for any dimensions we will then move on to outline the key
questions that any theory of entanglement will aim to answer when
it addresses the resource character of quantum entanglement. Some
major results and definitions from the theory of entanglement for
finite dimensional systems will be presented. From there we will
move on to prepare the ground for the study of the entanglement
properties of infinite-dimensional systems by introducing some
basic notions that are helpful for the description of states and
dynamics. Then we will exhibit some of the problems that occur
when one attempts to study entanglement of infinite-dimensional
systems without any restrictions. This will serve as a motivation
to restrict entanglement theory to so-called Gaussian states.
Again, basic results are presented; we will then ask the
experimentally motivated question of the manipulation of Gaussian
states by Gaussian operations. We will outline results that have
been obtained on this question, including a no-go theorem
concerning entanglement distillation of Gaussian states and
Gaussian operations, and results on general entanglement
manipulation for Gaussian states. In the conclusions we will
outline further developments and applications of the theory of
Gaussian entanglement in other areas of quantum information
science.

\section{Basic notions of entanglement}
The power of entanglement manifests itself particularly clearly in
quantum communication tasks that have to be performed over large
distances. Such tasks will typically require that, say, two
parties involved in the communication protocol establish a shared
entangled state of two quantum particles between them. This may
for example be accomplished by one party preparing two entangled
particles locally followed by the transmission of one of these
particles through a physical channel such as an optical fiber to
the other party. For large distances this task is complicated by
the unavoidable presence of unwanted interactions with the
environment causing decoherence and dissipation that will tend to
destroy entanglement. The two distant parties  that wish to create
a shared entangled state between them will therefore generally end
up with a partially entangled mixed state. Consequently, in order
to be able to implement an quantum communication protocol, they
will need to be able to 'repair' the partially entanglement state.
This aim will be hampered by the fact that the parties are
spatially separated, as this implies that they can only implement
general quantum operations in their respective local laboratories
and coordinate their respective actions by classical
communication. Global quantum operations affecting particles in
separate laboratories are not available to them. The set of local
operations and classical communication is usually abbreviated LOCC
and forms the basis for much of the study of quantum entanglement,
though more general classes of operations can also be relevant
\cite{Rains 01,Audenaert PE 03}. The very notion of entanglement
is actually tied to the set of LOCC: a state is called
disentangled if it can be prepared using LOCC only. Indeed, the
most general two-party state that can be generated from a product
state employing LOCC is a mixture of product states of the form
\begin{equation}
    \rho = \sum_{i} p_i \left(\rho_i^A \otimes \rho_i^B\right),
    \label{separable}
\end{equation}
where the tensor product refers to the parties $A$ and $B$. Such a
state, usually called a separable state, may exhibit correlations,
but they are not of genuinely quantum mechanical nature as, again,
this state may be created using LOCC \cite{Werner 89}. A separable
state allows for a description in terms of a local classical
model. Any state that cannot be cast into the form Eq.\
(\ref{separable}) (or appropriately approximated by such a state
for infinite dimensional systems) will be called entangled.
The development of a theory of entanglement, or in other words the
study of what can and cannot be achieved under LOCC, is a key
concern in the field. Such a theory will generally aim to answer
the following three central questions:

\subsection{Characterize}
While the definition of separability is easy
to formulate, it is very difficult to decide in practice whether a
given state  is separable or not. Following Eq.\
(\ref{separable}), in order to show that a state is separable, it
appears that one has to construct explicitly a decomposition of
the state into tensor products. This is a very difficult and
potentially lengthy task especially for high dimensional systems.
 For low dimensional systems, however, the separability
question can be decided in a different and more efficient way
using the theory of positive but not completely positive maps.  In
fact, a simple necessary and sufficient criterion for the
separability of a quantum state can be based on the properties of
the transposition and its application on a single sub-system.
Clearly, the transposition is a positive map in the sense that it
maps any positive operator onto a positive operator, i.e., if
$\rho$ is positive then so is $\rho^T$. The same then applies when
the transposition is applied to one sub-system, say system $B$, of
a separable state, because
\begin{equation}
    \rho^{T_B} = \sum_{i} p_i \left(\rho_i^A \otimes
    \left(\rho_i^B\right)^T\right)
\end{equation}
is again a valid state. However, when we apply this so-called
partial transposition to an inseparable state, then there is no
guarantee that the result is again a positive operator, i.e., a
physical state. Indeed, when applying the partial transposition to
the state with state vector $|\psi\rangle=(|01\rangle -
|10\rangle)/\sqrt{2}$ (with respect to the basis
$\{|00\rangle,|01\rangle,|10\rangle,|11\rangle \}$), we find
$|\psi\rangle\langle\psi|^{T_B} = \frac{1}{2}(|01\rangle\langle
01|+|10\rangle\langle 10|-|00\rangle\langle 11|-|11\rangle\langle
00|)$, which is evidently not positive any more. This observation
gave rise to the conjecture that a state is inseparable exactly if
the partial transpose is not a positive operator. It has indeed
been proven that this conjecture is correct for systems composed
of two qubits or a qubit and a qutrit \cite{Horodecki HH 96}. It
should be noted that the partial transposition on qubits
corresponds to time reversal. This observation already suggests
the appropriate application of the partial transposition criterion
for infinite dimensional systems. In systems with larger dimension
that six the situation is fairly different, and the question of
separability is far more complicated: one encounters for example
inseparable states whose partial transposition is nevertheless
positive again -- a phenomenon commonly referred to as bound
entanglement \cite{Horodecki 97,Bruss CHHKLS 02}. While there is a
systematic strategy for deciding the separability of a state
\cite{Doherty PS 03}, it is known that the problem has a
computational complexity that lies in NP-HARD \cite{Gurvits 02}.

\subsection{Manipulate}
A natural next question would be: Given a
quantum state of which one knows that it is entangled, how can it
be manipulated by LOCC? One may envision, for example, the
situation that an experimental procedure generates a particular
state $\rho$, but that a state $\rho'$ different from $\rho$ is
actually required. Is it possible to achieve the transformation
$\rho\longrightarrow\rho'$ employing LOCC only? In general, this
is a extraordinarily difficult question to answer, but some
results have been achieved for finite dimensional systems. Most
progress has been made for pure quantum states for which the basic
structures are known. There are two regimes which may fruitfully
be considered. Firstly, one may study the situation where a single
copy of a quantum state is held by two parties (the multiple copy
case is a special case of this setting). In this case, the
mathematical structure underlying the manipulation of quantum
states by LOCC is provided by the theory of majorization, and both
necessary and sufficient conditions for the interconversion
between two states are known, together with protocols that achieve
the task \cite{Nielsen 99,Vidal 99,Jonathan P 99a}. Let us write
the initial and final state vectors as $|\psi_1\rangle$ and
$|\psi_2\rangle$ in their Schmidt-basis,
\begin{eqnarray}
    \left| \psi_1 \right\rangle = \sum^{n}_{i=1}
    \sqrt{\alpha_i}\left|i_A \right\rangle \left|i_B
    \right\rangle,\;\;\;\;
    \left| \psi_2 \right\rangle = \sum^{m}_{i=1}\sqrt{\alpha'_i}
    \left|i_A' \right\rangle \left|i_B' \right\rangle ,
\end{eqnarray}
where $n$ denotes the dimension of each of the quantum systems.
We can take the Schmidt coefficients to be given in decreasing
order, i.e., $\alpha_1\ge\alpha_2\ge \ldots\ge \alpha_n$ and
$\alpha'_1\ge\alpha'_2\ge \ldots\ge \alpha'_n$. The question of
the interconvertability between the states can then be
decided from the knowledge of the real Schmidt coefficients only.
One finds that a LOCC transformation that converts
$\left|\psi_1\right\rangle $ to $\left| \psi_2 \right\rangle $
with unit probability exists if and only if the $\left\{\alpha_i
\right\}$ are  majorized \cite{Bhatia 97} by
$\left\{\alpha'_i\right\}$, that is, exactly if for all $1\leq
l\leq n$
\begin{equation}\label{majorization}
\sum_{i=1}^{l}\alpha _{i}\leq \sum_{i=1}^{l}\alpha _{i}^{\prime }.
\end{equation}
Various refinements of this result have been found  that provide
the largest success probabilities for the interconversion between
two states by LOCC together with the optimal protocol where a
deterministic interconversion is not possible \cite{Vidal
99,Jonathan P 99a}. These results allow in principle to decide any
question concerning the LOCC interconversion of pure state by
employing techniques from linear programming \cite{Jonathan P
99a}. Although the basic mathematical structure is well
understood, surprising and not yet fully understood effects such
as entanglement catalysis are still possible \cite{Jonathan P
99b}.

Secondly, further to the study of the LOCC transformation of
single copies, a fruitful area is the study of the asymptotic
regime ie the limit of a very large number of identical quantum
states. While this may be viewed as a limiting case of the above
theorem, it has indeed predated it. It turns out that in this
case, a single number, the entropy of entanglement, will determine
what is possible and what is not. The entropy of entanglement for
a pure state with state vector $|\psi\rangle$ is defined as
\begin{equation} \label{entropyofentanglement}
    E(|\psi\rangle\langle \psi|) = (S\circ \mbox{tr}_B)
    (|\psi\rangle\langle\psi|) \; .
\end{equation}
where $S$ denotes the von-Neumann entropy defined as
$S(\rho)=-\mbox{tr}[ \rho\log_2 \rho]$ and $\mbox{tr}_B$ denotes
the partial trace over subsystem B. With this quantity, we find
that in the asymptotic limit of a large number of identically
prepared systems, the transformation from $N$ copies of
$|\psi_1\rangle\langle \psi_1|$ under LOCC to
$M=N E(|\psi_1\rangle\langle \psi_1|)/E(\psi_2\rangle\langle
\psi_2|)$
copies of $|\psi_2\rangle\langle \psi_{2}|$ will be possible with
fidelity that approaches unity in the limit of $N$ approaching
infinity.

\subsection{Quantify}
For mixed states no comparable
set of theorems is available. Indeed, even very simple questions
concerning the interconvertability of states cannot be provided
with necessary and sufficient conditions for their availability.
The quest for such conditions together with the questions of the
best efficiencies for procedures of salient interest that are
available has led to the development of the theory of the
quantification of entanglement. An entanglement measure grasps
this notion of quantifying entanglement: it is a mathematical
quantity that captures the essential properties that we would
associate with the amount of entanglement and  that is ideally
related to some operational procedure. Any function $E$ mapping
the state space on positive real numbers that deserves the name
entanglement measure should satisfy the subsequent list of
requirements: \cite{Vedral P 98,Donald HR 01}
\begin{enumerate}
    \item $E(\rho)$ vanishes if the state $\rho$ is separable.
    \item $E$ is invariant under local unitary
    transformations, i.e., local basis changes.
    \item $E$ does not increase on average
    under LOCC, i.e.,
    \begin{equation}
        E(\rho) \ge \sum_{i} p_i E(\rho_i)
    \end{equation}
    where in a LOCC the state $\rho_i$ with label $i$
    is obtained with probability $p_i$.
    \item  For pure state $|\psi\rangle\langle\psi|$ the measure
    reduces to the entropy of entanglement
    \begin{equation}
        E(|\psi\rangle\langle\psi|) = (S\circ \mbox{tr}_B)
    (|\psi\rangle\langle\psi|) \; .
    \end{equation}
\end{enumerate}
A function $E$ mapping state space on positive numbers that
satisfies the first three conditions is called an entanglement
monotone while an entanglement measure also satisfies condition
(4). Often, convexity of $E$ or asymptotic continuity are taken as
additional reasonable properties (see, e.g., Ref.\ \cite{Donald HR
01}). In fact, conditions (1--3) together with asymptotic
continuity uniquely specify the entanglement measure to be the
entropy of entanglement on pure states (4), up to a real number
that is merely a scaling of the quantity \cite{Donald HR 01}.

These conditions alone, yet, do not uniquely specify a measure of
entanglement on mixed states. The extremal measures
\cite{Horodecki HH 00} giving bounds for all others are the
distillable entanglement \cite{Rains 01,Bennett DSW 96},
quantifying the degree to which maximally entangled states can be
extracted under LOCC from a given supply of identically prepared
states, and the entanglement cost \cite{Bennett DSW 96,Hayden H
00}, quantifying the resources in terms of maximally entangled
states necessary in a preparation procedure of a state under LOCC.
The entanglement cost is in fact nothing but the asymptotic
version of the entanglement of formation \cite{Bennett DSW
96,Wootters 98}. The relative entropy of entanglement quantifies
to which extent a given state can be distinguished from a
separable state \cite{Vedral P 98,Vedral PRK 97,Plenio VP 00}. In
its asymptotic version \cite{Audenaert EJPVD 01} it provides
moreover a tight upper bound for distillable entanglement.
Practically most straightforward to use is probably the negativity
\cite{Zyczkowski HSL 98,Eisert P 99}, which is known to be an
entanglement monotone \cite{PhD,Vidal W 02}, and whose logarithm
(up to a constant) has an interpretation as an asymptotic
entanglement cost \cite{Audenaert PE 03}. The very recently
developed 'squashed' entanglement \cite{Christandl W 03} is a
convex entanglement monotone that is moreover additive, a property
that it shares with the relative entropy of entanglement with
reversed arguments \cite{Eisert AP 03}, but it remains finite on
pure states. The same set of questions namely, characterization,
manipulation and quantification will have to be addressed in the
development of a theory of entanglement. In the remainder of this
article we will outline what has been achieved in this respect in
the infinite dimensional setting.

\section{The transition from finite to infinite dimensions}

All the above results concern bi-partite entanglement for systems
comprising finite dimensional sub-systems such as qubits or
qutrits. More recently, however, considerable effort has been
directed towards the study of infinite dimensional quantum systems
such as the photon number degree of freedom of light modes
\cite{Wu KHW 86,Bowen SLR 03,Furusawa SBFKP 98,Loock BK 00} or
nano-mechanical harmonic oscillators \cite{Roukes 01}, or the
state of cold atomic gases \cite{Schori SP 02}. At first sight,
one would expect the theory of entanglement to become
extraordinarily complicated due to the fact that now the Hilbert
space is no longer finite-dimensional. Indeed, without imposing
certain restrictions on the set of states under consideration, one
even loses such elementary properties as the continuity of
entanglement and its measures \cite{Eisert SP 02a} and questions
such as distillability become difficult to answer \cite{Parker BP
00}. In this subsection we will demonstrate some of the
counterintuitive properties \cite{Eisert SP 02a,Keyl SW 03,Clifton
HK 00} of entanglement measures in the infinite dimensional
setting and use them to motivate possible restrictions on the set
of states that one wishes to consider.

To clarify this issue, let us again consider the joint system to
be bi-partite, consisting of parts labeled $A$ and $B$, each of
which having a finite number of degrees of freedom. Both ${\cal
H}_A$ and $ {\cal H}_B$ of the joint Hilbert space ${\cal H}={\cal
H}_A\otimes {\cal H}_B$ are taken to be infinite-dimensional, as
in case of two modes of light. We will consider trace-norm
continuity, where the trace norm $\|.\|_1$ is defined as
$\|A\|_1=\mbox{tr}[| A|]= \mbox{tr}[(A^\dagger A)^{1/2}]$ for
trace-class operators $A$ \cite{Commentonchoiceofnorm}. It is now
quite straightforward to see that the entropy of entanglement
(\ref{entropyofentanglement}) is no longer trace-norm continuous.
The following sequence of states exemplifies this property of the
entropy of entanglement: the entropy of entanglement may be very
different from zero or even infinite for states that are
arbitrarily close to pure product states.
In this example, let $\sigma_0=|\psi_0\rangle\langle\psi_0|$ with
$|\psi_0\rangle = |\phi_{A}^{(0)}\rangle\otimes
|\phi_{B}^{(0)}\rangle$, and define $\{\sigma_k\}_{k=1}^{\infty}$ as
a sequence of pure states $\sigma_k=|\psi_k\rangle\langle\psi_k|$
defined by
\begin{eqnarray}
    |\psi_k\rangle &=&
    (1-\varepsilon_k)^{1/2}
    |\psi_0\rangle\nonumber\\
	 &+&  \left( \varepsilon_k/k\right)^{1/2}
    \sum_{n=1}^k |\phi_{A}^{(n)}\rangle\otimes |\phi_{B}^{(n)}\rangle,
\end{eqnarray}
 and $\varepsilon_k = 1/ \log (k)^2$. Here,
$\{|\phi_A^{(n)}\rangle: n\in{\mathbbm{N}}_0\}$ and
$\{|\phi_B^{(n)}\rangle : n\in{\mathbbm{N}}_0\}$ are orthonormal
basis that are dense in ${\cal H}_A$ and ${\cal H}_B$,
respectively. The sequence of states $\{\sigma_k\}_{k=1}^{\infty}$
in fact converges to $\sigma_0$ in trace-norm, i.e.,
$\lim_{k\rightarrow \infty} \| \sigma_k - \sigma_0\|_1=0$ while
$\lim_{k\rightarrow \infty}E(\sigma_k)=\infty$. Obviously, $E$ is
not continuous around the state $\sigma_0$. In fact, the states
with infinite von-Neumann entropy is trace-norm dense in state
space \cite{Wehrl 78}. As can readily be verified,
 the set of pure states with
infinite entropy of entanglement is also dense in the set of all
pure states, and hence, the entropy of entanglement is almost
everywhere infinite. That is to say, the assignment of a value
which represents the degree of entanglement to an
infinite-dimensional quantum system is not entirely unambiguous.

This apparent counterintuitive pathology can nevertheless be tamed
to a  significant degree. The key observation here is
concerned with the mean energy of the states. Let $H=H_A \otimes
{\id} + {\id} \otimes H_B$ be the Hamiltonian of the bi-partite
quantum system. It may appear to some extent unusual in the field
of quantum information theory to refer to the Hamiltonians of the
actual physical systems carrying the quantum states. We do not
require much from $H$ at all, however, only that for any finite
temperature $T$ we have
\begin{equation}
    {\mbox{tr}}[e^{-\beta H}]<\infty
\end{equation}
for all $\beta>0$ where $\beta=1/T$. This is a very natural demand
on the spectrum on $H$. It merely means that there can be no
limiting points in the spectrum of $H$: What it physically implies
is that the Gibbs state, the state of the canonical ensemble,
exists. Systems such as the photon number degree of freedom of
field modes of light, have this property. What are the
implications for the discontinuity of the entropy of entanglement?
We observe with $H_{A}= \sum_{k=0}^{\infty} k
|\phi_{A}^{(k)}\rangle\langle \phi_{A}^{(k)}|$ and $H_{B}=
\sum_{k=0}^{\infty} k |\phi_{B}^{(k)}\rangle\langle
\phi_{B}^{(k)}|$ that in the above example the mean energy of the
sequence of states diverges,
\begin{eqnarray}
       \lim_{k\rightarrow\infty}
       \mbox{tr}[\sigma_k H] &=& \lim_{k\rightarrow\infty}\sum_{n=1}^{k}
       \frac{1}{\log(n)^{2}}= \infty.
\end{eqnarray}
This divergence of the mean energy is actually generic for
sequences of states with divergent entropy of entanglement
\cite{Eisert SP 02a}, and is hence not an accident. This
observation suggests how to tame the unwieldy infinities. After
all, the energy that can be invested in the preparation of a state
by means of physical devices is in all instances limited. This
does not mean that the resulting states have a finite-dimensional
carrier. It only means that their {\it mean} energy must be
bounded. In fact, it turns out that for the set of states with
mean energy which is bounded from above by a positive number
$E_{max}$,
\begin{equation}\label{energy}
     \{ \rho\in {\cal S}({\cal H}) \,:\, \mbox{tr}[\rho H] \leq {E_{max}}\},
\end{equation}
the entropy of entanglement regains its (trace-norm) continuity
\cite{Eisert SP 02a,Wehrl 78}. Here, ${\cal S}({\cal H})$ denotes
the state space, i.e., the set of all density matrices, which is
the set of all positive normalized trace-class operators. The set
Eq.\ (\ref{energy}) is technically speaking nowhere dense in state
space as $H$ is an unbounded operator, but it is a reasonable
subset of the state space: it simply reflects the practically
natural requirement that the mean energy is bounded from above.

Moreover,  restricting the mean energy of the states under
consideration recovers  a number of asymptotic continuity
properties of the entropy of entanglement. This strengthens the
interpretation as the entanglement cost  and the distillable
entanglement also in the infinite-dimensional setting \cite{Eisert
SP 02a}.  Similar conclusions can be reached for the entanglement
of formation and the relative entropy of entanglement \cite{Eisert
SP 02a}. In a nutshell, one may say that on subsets of state space
corresponding to bounded mean energy, entanglement measures often
regain trace-norm continuity. Nevertheless, one has to keep in
mind that the actual minimization problems involved in the
evaluation of meaningful entanglement measures are often not
feasible in the infinite-dimensional setting. This may again be
taken as a very discouraging observation. However, in many
situations of practical interest, where the semantics of the term
practical may range from 'meaningful in mathematical physics' to
'preparable in a quantum optical experiment', it turns out that
one does not encounter all possible states from state space.
Instead, a most relevant class of quantum states can be described
in very simple terms using (small) finite-dimensional matrices
only, without the need of a description that is overburdened with
the technicalities of infinite-dimensional Hilbert spaces: this is
the set of Gaussian quantum states. This set of states that is of
utmost practical importance will be dealt with in most of the
remainder of the present article.

\section{Entanglement properties of Gaussian states}

Gaussian quantum states play a key role in several fields of
theoretical physics. In quantum optics, for example, they are
often encountered as states of field modes of light, for reasons
that will be elaborated on below. Ground states of systems with
canonical coordinates (position and momentum) where the
Hamiltonian is quadratic in the positions and momenta are also
Gaussian states \cite{Audenaert EPW 02}. The term has been coined
because the defining property is that the characteristic function
associated with the state is a Gaussian in phase space. For such
Gaussian states the theory of entanglement as well as a framework
of how these states may be manipulated is  well developed. Typical
questions  in a theory of entanglement, e.g., concerning  the
separability of given states, or the interconvertability of pairs
of states under local operations can often be answered in full.
This is true even in cases where the finite dimensional
equivalent, whenever such an equivalent can be formulated, remains
unsolved. One key reason for these successes is the fact that
Gaussian states are completely specified by their first and second
moments so that questions concerning properties of Gaussian states
can be translated into properties of (comparatively  small)
finite-dimensional matrices. Therefore the instruments of matrix
theory \cite{Bhatia 97}, so useful in the study of entanglement
for finite-dimensional systems such as qubits and qutrits, once
more become a useful tool.

The systems that will subsequently be discussed are quantum
systems with $n$ canonical degrees of freedom. These could
represent $n$ harmonic oscillators, or $n$ field modes of light.
The canonical commutation relations (CCR) between the $2n$
canonical self-adjoint operators corresponding to position and
momentum of such a system with $n$ degrees of freedom, may be
written in a particularly convenient form employing the row
vector,
\begin{equation}
    {{O}}=(O_1,\ldots,O_{2n})^T=
    (X_1,P_1,\ldots,X_{n},P_{n})^T.
\end{equation}
 In terms of the familiar creation and annihilation operators
of the modes choosing $\hbar=1$, $X_n$ and $P_n$ can be expressed
as
\begin{equation}
    X_n=(a_n+ a_n^\dagger)/\sqrt{2},\,\,
    P_n=- i (a_n- a_n^\dagger)/\sqrt{2}.
\end{equation}
Then the canonical commutation relations (CCR) take the form
\begin{equation}
    [O_j,O_k]=i \sigma_{j,k} \; ,
\end{equation}
where the skew-symmetric block diagonal real $2n\times 2n$-matrix
$\sigma$ given by
\begin{equation}
    \sigma =\bigoplus_{j=1}^{n}
    \left[\begin{array}{cc}
    0 & 1 \\
    -1 & 0 \\
    \end{array}
    \right],
\end{equation}
is the so-called symplectic matrix. The phase space, isomorphic to
${\mathbbm{R}}^{2n}$, then becomes what is known as a symplectic
vector space, equipped with the scalar product corresponding to
this symplectic matrix. Instead of referring to states, i.e.,
density operators, one may equivalently refer to functions that
are defined on phase space. There are many common choices of such
functions in phase space, such as the Wigner function, the
$Q$-function or the $P$-function, to name just a few
\cite{Schleich 01}. Each of them is favorable in a particular
physical context. For later purposes it is most convenient to
introduce the characteristic function, which is the Fourier
transform of the Wigner function. Using the Weyl operator
\begin{equation}
    W_{\xi} = e^{i  \xi^{T} \sigma O}
\end{equation}
for $\xi\in{\mathbbm{R}}^{2n}$, we define the (Wigner-)
characteristic function as
\begin{equation}
    \chi_{\rho}(\xi) = \mbox{tr}[\rho W_{\xi}].
\end{equation}
In quantum optics, the Weyl operator is typically referred to as
phase space displacement operator or Glauber operator, but with a
different convention concerning its arguments. For a single mode,
let the complex number $\alpha$ be defined as $\alpha=-(\xi_1 + i
\xi_2)/\sqrt{2},\alpha^\ast =-(\xi_1 + i \xi_2)/\sqrt{2}$,
then the phase space displacement operator $D_\alpha$ of quantum
optics \cite{Walls M 94} is most commonly taken as
$D_\alpha=W_{\xi}$. Each characteristic function is uniquely
associated with a state, and they are related with each other via
a Fourier-Weyl relation. The state $\rho$ can be obtained from its
characteristic function according to
\begin{equation}
    \rho = \frac{1}{(2\pi)^n}
        \int d^{2n}\xi
    \chi_{\rho}(-\xi) W_{\xi} .
\end{equation}
In turn, the Wigner function as commonly used in quantum optics is
related to the characteristic function via a Fourier transform,
i.e.,
\begin{equation}
    {\cal W}(\xi)=\frac{1}{(2\pi)^{2n}}
    \int
    d^{2n} \zeta
    e^{i \xi^{T}
    \sigma \zeta} \chi(\zeta) .
\end{equation}
Gaussian states are, as mentioned before, defined through their
property that the characteristic function is a Gaussian function
in phase space \cite{Simon SM 87}, i.e.,
\begin{equation}
    \chi_{\rho}(\xi) =
    \chi_{\rho}(0) e^{-\frac{1}{4}\xi^T \Gamma \xi + D^T
    \xi},
\end{equation}
where $\Gamma$ is a $2n\times 2n$-matrix and
$D\in{\mathbbm{R}}^{2n}$ is a vector. As a consequence, a Gaussian
characteristic function can be characterized via its first and
second moments only, such that a Gaussian state of $n$ modes
requires only $2n^2+n$ real parameters for its full description,
which is polynomial rather than  exponential in $n$. The first
moments form a vector, the displacement vector
$d\in{\mathbbm{R}}^{2n}$, where
\begin{equation}
    d_j= \langle O_j\rangle_\rho= \mbox{tr}[O_j \rho ],
\end{equation}
$j=1,...,2n$. They are the expectation values of the canonical
coordinates, and are linked to the above $D$ by $D= \sigma d$.
They can be made zero by means of a translation in phase space of
individual oscillators. As a consequence the first moments do not
carry any information about the entanglement properties of the
state. The second moments are embodied in the real symmetric
$2n\times 2n$ covariance matrix $\gamma$ defined as
\begin{eqnarray}
    \gamma_{j,k}
    &=&
    2 \mbox{Re} \, \mbox{tr}\left[
    \rho \left(O_j-\langle O_j\rangle_{\rho} \right)
    \left(O_k-\langle O_k\rangle_{\rho} \right)
    \right] \, .
\end{eqnarray}
With this convention, the covariance matrix of the $n$-mode vacuum
is simply $\id_{2n}$. Again, the link to the above matrix $\Gamma$
is $\Gamma= \sigma^{T}\gamma \sigma$. Clearly, not any real
symmetric $2n\times 2n$-matrix can be a legitimate covariance of a
quantum state: states must respect the Heisenberg uncertainty
relation. In terms of the second moments the latter can be phrased
in compact form as the matrix inequality
\begin{equation}\label{res}
    \gamma + i \sigma\geq 0 \, .
\end{equation}
In turn, for any real symmetric matrix $\gamma$ satisfying the
uncertainty principle (\ref{res}) there exists a Gaussian state
the second moments of which are nothing but $\gamma$. So Eq.\
(\ref{res}) implies the only restriction on legitimate covariance
matrices of Gaussian quantum states.

This observation has quite significant implications concerning the
question of separability of two-mode Gaussian states shared by two
parties: it has earlier been pointed out that partial
transposition is a positive, but not completely positive map.
Hence, partial transposition must map separable states onto
separable states, but there exist states for which the partial
transpose is no longer positive. In fact, for bipartite qubit
systems the positivity of the partial transpose is a necessary and
sufficient criterion for separability. A necessary condition for
separability of Gaussian states can be formulated immediately,
once it is understood how partial transposition is reflected on
the level of covariance matrices. Indeed, it has been pointed out
earlier that the partial transposition on qubits is in fact
time-reversal. Time reversal in a system with canonical degrees of
freedom is characterized by the transformation that leaves the
positions invariant but reverses all momenta,
\begin{equation}
   X \longmapsto X \;\;\;\; P \longmapsto -P.
\end{equation}
Let us now consider a system made up of $2n$ oscillators, where
$n$ are held by each party. Applying the time reversal operation
to the $n$ oscillators held by one of the parties, the covariance
matrix will be transformed to a real symmetric matrix ${\tilde
\gamma}$ given by
  $  {\tilde \gamma}= F\gamma  F$,
with
\begin{equation}\label{f}
    F= {\mathbbm{1}}_{2n}\oplus
    \bigoplus_{i=1}^n \left[\begin{array}{cc} 1 & 0 \\ 0 & -1 \end{array}\right] .
\end{equation}
The matrix $\tilde \gamma$ is the matrix collecting the second
moments of the partial transpose $\rho^{T_B}$ of $\rho$. As the
positivity of $\rho^{T_B}$ is equivalent to $\tilde \gamma$
satisfying the Heisenberg uncertainty principle, one may now
merely check whether
\begin{equation}\label{crit}
    {\tilde \gamma} + i \sigma\geq 0 \, .
\end{equation}
If ${\tilde \gamma} + i \sigma$ is not positive, then the state
 $\rho$  must in fact be entangled. For two-mode
systems, one mode held by each party, the criterion (\ref{crit})
is both necessary and sufficient for separability of the Gaussian
state. One direction of the proof has been exemplified above, the
converse has been proven in Refs.\ \cite{Simon 00,Duan GCZ 00}.
Hence, the two-mode Gaussian state with covariance matrix $\gamma$
is separable exactly if the covariance matrix ${\tilde \gamma}$
obtained via partial time-reversal in one sub-system also
represents a valid physical state, i.e., if ${\tilde
\gamma}+i\sigma\geq 0$.

It is remarkable that the above theorem can be extended
considerably. In fact, for a bi-partite system the partial
time-reversal also provides a necessary and sufficient criterion
for the distillability of a bi-partite Gaussian state of two
party holds one mode and the other party holds $N$ modes,
positivity of the partial transpose is still a necessary and
sufficient criterion for separability \cite{Werner W 01}. This
result together with the discovery of bound entanglement in
Gaussian systems rely on an alternative criterion for the
separability of Gaussian states. In fact, a Gaussian state
represented by the covariance matrix $\gamma$ is separable if and
only if there exist covariance matrices $\gamma_A$ and $\gamma_B$
such that
\begin{equation}\label{generalcrit}
    \gamma\geq \gamma_A\oplus \gamma_B.
\end{equation}
This formulation is a very useful tool to prove the validity of
statements, but does not serve as a practical criterion for
separability, as it is generally not straightforward to find the
matrices $\gamma_A$ and $\gamma_B$ (but note that
this may be checked in form of a semi-definite program). 
In general, given two parties,
one holding $M$ and the other $N$ oscillators, both the question
of separability \cite{Giedke DCZ 01,Giedke KLC 01} and
distillability can be decided efficiently, although the questions
of distillability requires a more sophisticated approach
\cite{Giedke DCZ 01}.

These results are very encouraging but they come with one caveat.
While the states have been restricted to be Gaussian, no such
restriction has been  applied to the set of  allowed operations.
The result concerning the distillability for example only holds
when arbitrary operations in infinite-dimensional systems are
permitted, even practically very complicated ones such as the
distinction between $999$ and $1000$ photons in field modes of
light. In the light of this observation, one should not forget
that much of the significance of the Gaussian states stems from
the fact that the Gaussian operations are so important. This is by
no means a tautology: The Gaussian operations are those operations
that change the state, but preserve the Gaussian character of any
Gaussian input state. This set of operations is singled out from
both a theoretical perspective, but most significantly also from
the perspective of practical experimental implementation. A most
relevant question is then what tasks can be accomplished with
Gaussian states, but not under {\it all}  physically available
 operations, but under Gaussian operations only. This
observation motivates the approach described in the next section,
namely the investigation of what can and cannot be achieved when
one restricts attention to Gaussian states and the so-called
Gaussian operations.

\section{Characterizing Gaussian operations}
Before we can begin the development of the theory of entanglement
of Gaussian states under Gaussian operations, we need to
characterize both Gaussian states (already done in the earlier
sections) and Gaussian operations. The most straightforward
definition of Gaussian operations simply states that an operation
is Gaussian exactly if it will map every  Gaussian input state
onto a Gaussian output state. The remainder of this section will
present more practically useful characterizations of Gaussian
operations and in particular it will show that Gaussian operations
correspond exactly to those operations that can be implemented by
means of optical elements such as beam splitters, phase shifts and
squeezers together with homodyne measurements \cite{Giedke C
02,Eisert SP 02b,Fiuraszek 02} -- all operations that are in
principle experimentally accessible with present technology
\cite{Leonhardt}.

Let us start with Gaussian unitary transformations. The most
general real linear transformation S which implements the mapping
    $S:O\longmapsto O'=S O$
will have to preserve the canonical commutation relations
$[O'_j,O'_k]=i\sigma_{jk}\id \,.$
This is the case exactly if the linear map $S$ satisfies
\begin{equation}
    S\sigma S^T=\sigma \, .
\end{equation}
The set of real $2n\times 2n$ matrices $S$ obeying this condition
form the so-called real symplectic group $Sp(2n,\mathbb{R})$
\cite{Arvind DMS 95}, whose elements are called symplectic or
canonical transformations. To any symplectic transformation $S$
also $S^T,S^{-1},-S$ are symplectic. The inverse of $S$ is given
by
$
    S^{-1}= \sigma S^T \sigma^{-1}$
and the determinant of every symplectic matrix is $\det[ S] =1$
\cite{Arvind DMS 95,Simon SM 87}.
Every symplectic transformation corresponds to a unitary operation
acting on the corresponding state space. As a consequence of the
Stone-von Neumann theorem, given a real symplectic transformation
$S$ there exists a unique unitary transformation $U_S$ acting on
the state space such that the Weyl operators satisfy
 $U_S W_{\xi} U_S^\dagger = W_{S\xi} $
 for all $\xi\in {\mathbbm{R}}^{2}$.
%
%
On the level of covariance matrices $\gamma$ of an $n$-mode system
a symplectic transformation $S$ is reflected by a congruence
\begin{equation}
    \gamma \longmapsto S \gamma S^T .
\end{equation}
It is instructive to consider the generators of the Gaussian
unitary transformations. These infinitesimal generators can be
found by applying the condition that the canonical commutation
relations (on the level of Weyl operators) are preserved to an
infinitesimal Gaussian unitary transformation. We consider
 $   U = e^{-i\epsilon G} = \id - i\epsilon G + o(\epsilon^2)$,
for a real $\epsilon$ and arrive at the conclusion that the
generators $G$ have the form $G = \sum_{jk=1}^{2n} g_{jk} (O_jO_k
+ O_kO_j)/2$, which is a Hermitian quadratic expression in the
canonical operators (and thus in the annihilation/creation
operators). As a consequence, every Hamiltonian $G$ can in quantum
optical settings be implemented using beam-splitters, phase plates
and mirrors only. Conversely, every array of passive optical
elements is described by such a Hamiltonian.

A particularly important subset of all symplectic transformations
is formed by those $S\in Sp(2n,{\mathbbm{R}})$ that are moreover
orthogonal, i.e., $K(n) =Sp(2n, {\mathbbm{R}})\cap O(2n)$. $K(n)$
is again a group, and a compact one. The elements of $K(n)$
correspond to the passive transformations \cite{Arvind DMS 95}. In
optical systems, these particular Gaussian unitary operations are
those that can be realized using beam splitters and phase shifts,
but excluding squeezers. Such a passive transformation does not
change the eigenvalues of the covariance matrix and in an optical
systems does not change the total photon number. Hence, it can
never transform a state that is not squeezed into a squeezed one.
A state is called squeezed, if one of the eigenvalues of the
covariance matrix $\gamma$ is smaller than unity \cite{Simon SM
87}. In more physical terms this means that using passive
transformations only the state can be transformed into a state
such that the uncertainty with position is smaller than the
respective uncertainty of the vacuum state.
Any $S\in Sp(2n, {\mathbbm{R}})$ can be decomposed into
\begin{equation}
    S= K \bigoplus_{j=1}^n \left[
    \begin{array}{cc}
    d_j & 0\\
    0 & 1/d_j\\
    \end{array}
    \right]
    L,
\end{equation}
 with $K,L\in K(n)$ and $d_j\in{\mathbbm{R}}$. This
 is the Euler decomposition of symplectic transformations.
 Physically, this means that any
Gaussian unitary transformation can be realized as a (i) passive
transformation, a (ii) single-mode squeezing operation on each of
the $n$ modes, and a (iii) subsequent second passive
transformation.

Noisy channels, such as those corresponding to optical fibers in
optical systems, can only be modeled as an irreversible quantum
operation. A Gaussian channel \cite{Demoen VV 79,Holevo W
01,Eisert P 02a} is such an irreversible quantum operation,
    $\rho \longmapsto {\cal E}(\rho)$.
 It is reflected by a
transformation of the covariance matrix according to
\begin{equation}\label{channel}
    \gamma \longmapsto A \gamma A^T + G.
\end{equation}
where $A,G$ are $2n\times 2n$-matrices, and $G$ is moreover
symmetric. The complete positivity of the quantum operation is
reflected by the condition
\begin{equation}\label{cp}
    G + i \sigma - i A \sigma A^T \geq 0.
\end{equation}
Such channels emerge naturally if a system is coupled to degrees
of freedom that cannot be controlled. Often, this interaction is
of Gaussian nature, as the Hamiltonian is a polynomial of second
degree in the canonical coordinates \cite{Paris ISS 03}. As the
environment degrees of freedom cannot be accessed, one considers a
partial trace with respect to them. Optical fibers manifest
themselves for example as Gaussian channels of this type for field
modes \cite{Scheel KOW 00}. Then, the overall dilation is
reflected by a Gaussian completely positive map with a
transformation rule for the second moments according to Eq.\
(\ref{channel}). Note that there are no restrictions on the matrix
$A$: this means that if one aims to implement an (impossible)
operation with a transformation of the form $\gamma\longmapsto
A\gamma A^T$, Eq.\ (\ref{channel}) together with (\ref{cp}) show
the fundamental limitations imposed by the CCR on this operation.
This is relevant in the context of, for example, quantum
amplifiers, attenuators, or phase conjugating mirrors
approximately realising a time reversal operation. Also, quantum
cloning \cite{Lindblad 00,Cerf IR 00} and secret sharing \cite{Tyc
RS 03} have been formulated as a Gaussian channel`.

In turn, one may ask,  whether  it is true that any channel of
this form can be realized in three steps: (i) appending an
additional system prepared in a Gaussian state ('ancillae'), (ii)
a Gaussian unitary operation including all modes, and (iii) a
partial trace with respect to the environment ('discarding the
ancillae')?  The structure of such a dilation has been depicted in
Fig.\ \ref{figure4}. Clearly, any such operation is reflected by a
transformation of the form Eq.\ (\ref{channel}). The converse is
not quite true, and one might have to 'add classical Gaussian
noise', by displacing the state in phase space with a classical
Gaussian distribution. This is reflected by the addition of a
positive matrix to the covariance matrix.

Furthermore, one might after all perform selective measurements
which preserve the Gaussian character of the state. Let us assume
that $n$ modes of light are entangled with a distinguished $n+1$st
mode. If one projects the mode with label $n+1$ onto the vacuum
state with state vector $|0\rangle$, then the resulting state of
the remaining $n$ modes is still a Gaussian state. But what are
its second moments? If the original covariance matrix is of the
form
\begin{equation}
    \gamma =\left[
    \begin{array}{cc}
    A & C\\
    C^T & B
    \end{array}
    \right],
\end{equation}
where $A$ is an $2n\times 2n$-matrix and $B$ is a $2\times 2$
matrix corresponding to the distinguished mode, then the resulting
covariance matrix can be determined  by a Gaussian
integration  \cite{Eisert SP 02b}
\begin{equation}
    \gamma\longmapsto
    A-C (B+ {\id}_{2n})^{-1} C^T,
\end{equation}
which is often referred to as the Schur complement of the matrix
\begin{equation}
    \left[
    \begin{array}{cc}
    A & C\\
    C^T & B  + {\id}_{2n}
    \end{array}
    \right].
\end{equation}
Note that this transformation is not of the form as specified in
Eq.\ (\ref{channel}). Now, it follows how an ideal homodyne
detection is reflected on the level of second moments. The
resulting covariance matrix does not depend on the outcome of the
measurement \cite{Eisert SP 02b}. It is again a Schur complement
of a matrix. Indeed, with $\pi=\mbox{diag}(1,0)$ we have
\begin{equation}\label{homo}
    \gamma\longmapsto
    A-C (\pi B \pi)^{-1} C^T,
\end{equation}
where the inverse is replaced by the Moore-Penrose pseudo-inverse,
as the matrix is apparently no longer invertible. Eq.\
(\ref{homo}) gives the transformation law for the covariance
matrix of $n$ modes, after the mode with label $n+1$ has undergone
an ideal homodyne detection. In fact, any Gaussian operation can
be thought of as a concatenation from ingredients of this above
list \cite{Giedke C 02,Eisert SP 02b,Fiuraszek 02}, i.e., any
physical operation mapping all Gaussians onto Gaussians can be
thought of as a sequence of Gaussian channels (including Gaussian
unitary operations), together with homodyne detection. This notion
can be stated most clearly in terms of the isomorphism between
positive operators and completely positive maps \cite{Jamiolkowski
72}. For Gaussian operations this isomorphism \cite{Giedke C
02,Fiuraszek 02} leads to the following form of a mapping of the
general Gaussian operation, on the level of second moments,
\begin{eqnarray}
    \gamma\longmapsto
    \tilde\Gamma_1-\tilde\Gamma_{12}(\tilde\Gamma_2+\gamma)^{-1}
    \tilde\Gamma_{12}^T,
\end{eqnarray}
where we have denoted
\begin{equation}\label{capGamma}
    \Gamma = \left[ \begin{array}{cc} \Gamma_1&\Gamma_{12}\\
    \Gamma_{12}^T&\Gamma_2
    \end{array}\right] ,
\end{equation}
and $ \tilde\Gamma = F\Gamma F$, with $F$ being defined as in Eq.\
(\ref{f}) with appropriate size. All Gaussian completely positive
maps on $n$ modes are characterized by a $2n$-mode covariance
matrix $\Gamma\geq i\sigma_{2n}$ and a displacement vector $d\in
\mathbb{R}^{4n}$ \cite{Giedke C 02}. Equipped with the
transformation laws under these operations we may now turn to
assessing what state transformations can be done in the Gaussian
setting.

\section{Manipulation of Gaussian states under Gaussian operations}\label{s4}

In the last section we have presented useful characterizations of
Gaussian operations. These characterizations form the essential
tools for the study of the Gaussian local manipulation of Gaussian
states and play important roles in many proofs in this area. We
will now proceed with an exposition of known results in the area
that have been discovered in the past few years. It should be
noted however, that we will not proceed along the historical lines
of development. While the program of the study of Gaussian states
under Gaussian operations has originated in Ref.\ \cite{Eisert P
02a} for mixed Gaussian states, we will begin with an exposition
of the state of knowledge for pure states.


\subsection{Pure state entanglement}

\begin{figure}[hbt]
\centerline{\epsfxsize=\columnwidth\epsfbox{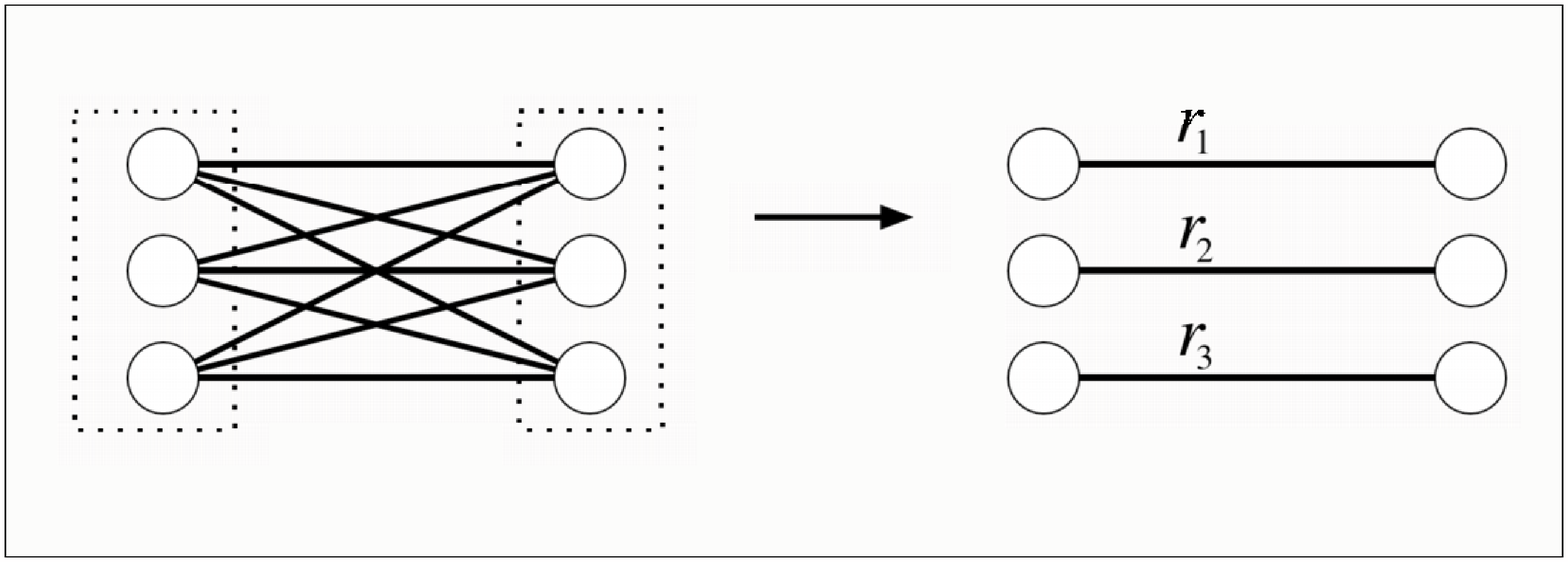}}
\vspace*{0.2cm} \caption{\label{figure3} Schmidt decomposition of
pure bi-partite Gaussian states.}
\end{figure}

As can be expected, the most
complete understanding of the behaviour of Gaussian states under
Gaussian operations has been reached for pure states. Indeed,
necessary and sufficient conditions for the possible state
transformations under local Gaussian operations (GLOCC) can be
given. Local means here that the applied Gaussian quantum
operations are applied locally, accompanied by classical
communication. This result crucially depends on a normal form that
can be obtained for any bipartite system of continuous variables
\cite{Holevo W 01,Giedke ECP 03,Botero R 03}. For any pure
Gaussian state with covariance matrix $\gamma$ of an $n\times
n$-mode system, there exist local symplectic transformations
$S_{A},S_{B}$ such that
\begin{eqnarray} \label{luf}
    &&(S_{A}\oplus S_{B}) \gamma
     (S_{A}\oplus S_{B})^{T}\\
	&=&
     \bigoplus_{k=1}^{n}
     \left[
    \begin{array}{cccc}
    \cosh(2r_{k})&0&\sinh(2r_{k})&0\\
    0&\cosh(2r_{k})&0&-\sinh(2r_{k})\\
    \sinh(2r_{k})&0&\cosh(2r_{k})&0\\0&-\sinh(2r_{k})&0&\cosh(2r_{k})
    \end{array}\right]\nonumber
\end{eqnarray}
with $r_{k}\in[0,\infty)$. In other words, by means of local
unitary Gaussian operations the pure Gaussian state described by
$\gamma$ can be transformed into a tensor product of two-mode
squeezed states
\begin{equation}\label{twomodesqueezed}
    |\psi_k\rangle = (1- (\tanh r_k/2)^2)^{1/2} 
	\sum_{n=0}^{\infty} (\tanh r_k/2)^n |n,n\rangle
\end{equation}
characterised by squeezing parameters $r_{k}$ (see Fig.\
\ref{figure3}).
As a consequence the non-local properties of a pure Gaussian state
of $n\times n$ modes can be characterised by a vector
\begin{equation}
    r=(r_{1},\ldots,r_{n})^{T}
\end{equation}
of two-mode squeezing parameters, which we will always assume to
be given in descending order. This vector will play a role closely
analogous to that of the Schmidt-coefficients in the majorization
criterion for pure state entanglement transformations under
general LOCC operations. The criterion for pure-state
transformations under GLOCC can now be stated in a very simple
manner in terms of this vector of two-mode squeezing parameters.
We write $   r\geq r'$ iff $r_{k}\geq r'_{k}$ for all
$k=1,\ldots,n$, in descending order. Then, the criterion for the
possibility of transforming the initial to the final state can be
expressed in an extraordinarily simple manner: we have that
\begin{equation}\label{MainCrit}
    \rho\longrightarrow \rho'\,\mbox{ under GLOCC,\, iff }r\geq r'.
\end{equation}

\begin{figure}[hbt]
\centerline{\epsfxsize=\columnwidth\epsfbox{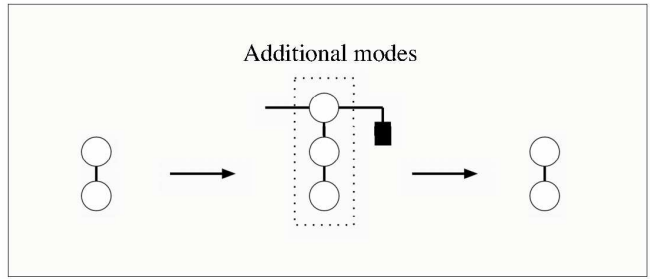} }
\vspace*{0.5cm} \caption{\label{figure4} Representation of a
Gaussian channel.}
\end{figure}

It should be noted that the criterion for pure state
transformations under Gaussian local operation has a simpler
structure than the criterion that determines which state
transformations are possible under general LOCC, not restricted to
Gaussian transformations. As already stated in earlier sections,
we have that
\begin{equation}\label{Majo}
    \rho\longrightarrow \rho'\,\mbox{ under LOCC iff}\;\;
    \lambda(\mbox{tr}_{A}[\rho])\prec
    \lambda(\mbox{tr}_{A}[\rho']).
\end{equation}
where  $\prec$ denotes the majorisation relation Eq.\
(\ref{majorization}). In other words, a pure state $\rho$ can be
transformed into another pure state $\rho'$ if and only if the
reduction to one part of the joint system is initially more mixed
than finally in the sense of majorisation theory.

This criterion for finite-dimensional systems can be immediately
carried over to the infinite-dimensional case, when appropriate
care is taken when extending the definition of LOCC
convertibility: one then writes $\sigma \longrightarrow \rho$
under LOCC if there exists a sequence of LOCC, such that the
images of the maps approximate $\rho$ in trace-norm. It is then
instructive to compare the restrictions imposed by the
majorisation criterion with the constraints under Gaussian LOCC.
To start with, under LOCC, catalysis of entanglement manipulation
is possible \cite{Jonathan P 99b,Eisert W 00}: there exist pure
states of finite-dimensional bi-partite quantum systems $\rho$,
$\rho'$ and $\omega$ such that
\begin{eqnarray}
    \rho& \not\longrightarrow& \rho'\,\,\,\mbox{ under LOCC, but }\nonumber\\
    \rho\otimes \omega & \longrightarrow &\rho'\otimes \omega
    \,\,\,\mbox{ under LOCC} \, .
\end{eqnarray}
The state $\omega$ serves as a ``catalyst'', as the entanglement
inherent in the state is not consumed in the course of the
transformation. Such an effect is not possible under GLOCC and
pure Gaussian states, as collective operations are never more
powerful  than  operations on pairs of quantum systems at
a time. This notably refers to entanglement transformations only,
as one can think of procedures estimating certain properties of
Gaussian states where collective operations are superior to
product operations. It should moreover be noted that the theorem
governing interconvertability under GLOCC also covers the
stochastic interconversion of pure Gaussian under GLOCC, as it has
been pointed out in the previous section that any Gaussian
transformation can be lifted to a trace-preserving operation. This
is again in contrast to the finite-dimensional case \cite{Vidal
99,Jonathan P 99a}.

It is also an immediate but remarkable consequence of the above
criterion that under Gaussian local operations with classical
communication, one cannot ``concentrate'' two-mode squeezing of
pure Gaussian states. In particular,
\begin{equation}
    \rho(r)^{\otimes n} \not\longrightarrow \rho(r') \otimes
    \rho(0)^{\otimes(n-1)}, \,\,\mbox{ under GLOCC},
\end{equation}
for any $n\in{\mathbb{N}}$ and any $r'>r$, where $\rho(r)$ denotes
a two-mode squeezed state with squeezing parameter
$r\in[0,\infty)$ . This is very much in contrast to the situation
when general operations are allowed for: for any $r\in[0,\infty)$
there exists a $r'>r$ such that
\begin{equation}
    \rho(r)^{\otimes 2} \longrightarrow \rho(r') \otimes \rho(0)
    \,\,\mbox{ under LOCC},
\end{equation}
as has been shown in Ref.\ \cite{Giedke ECP 03}. That is to say,
this criterion already points towards the impossibility of
distilling Gaussian states with Gaussian operations.

\subsection{Mixed states}

For mixed states, both the number of
parameters characterizing the state increases and the structure of
state transformations becomes far more involved. To start with, a
simple normal form such as the Schmidt decomposition or the vector
of two-mode squeezing parameters $r$ is not available. The
simplest normal form that can be obtained for two harmonic
oscillators is given by the Simon normal form for the covariance
matrix \cite{Simon 00}. If we are given a Gaussian state $\rho$ of
a two-mode system with covariance matrix $\gamma$, then there
exist symplectic transformations $S_{A},S_{B}\in
Sp(2,{\mathbb{R}})$ such that
\begin{equation}
    (S_{A}\oplus S_{B}) \gamma(S_{A}\oplus S_{B})^{T}=
    \left[
              \begin{array}{cccc}
               x_{1}&0&x_{3}&0\\
        0&x_{1}&0&x_{4}\\
                x_{3}&0&x_{2}&0\\0&x_{4}&0&x_{2}
              \end{array}\right].
\end{equation}
Further to the enlarged number of parameters characterizing the
state, also the amount of classical communication in general
Gaussian local operations cannot be bounded anymore as opposed to
the pure state case where one round of one-way communication is
sufficient. This latter fact has also been made use of in the
proof of the key statement of Ref.\ \cite{Giedke ECP 03}.
Nevertheless, under certain restrictions on the available
protocols and with the help of the Simon normal form it is
possible to obtain some statements concerning the
interconvertability of mixed states and one can find necessary and
sufficient criteria for a transformation to be possible. One such
restriction is that to local Gaussian channels as have been
discussed above: they can be realized by adding an ancilla system,
followed by a joint unitary transformation and the subsequent
discarding of the ancilla (see Fig.\ \ref{figure4} for
illustration).

For such local Gaussian channels, one may write
\begin{equation}
    \rho\longrightarrow \rho'\,\,\,\,\mbox{ under LOG}
\end{equation}
for Gaussian two-mode states $\rho$ and $\rho'$, if there exists
local Gaussian channels ${\cal E}_{A}$ and ${\cal E}_{B}$ such
that
\begin{equation}
    ({\cal E}_{A}\otimes {\cal E}_{B}) (\rho)= \rho'
\end{equation}
(LOG stands  for local Gaussian channel).
In fact, both necessary and sufficient conditions for the
interconvertability under LOG can then be proven. For details we
refer the reader to the literature, in particular Ref.\
\cite{Eisert P 02a} and for generalizations for example to a
three-party version of the above statement \cite{Wang LZ 02}.

\subsection{Distillation of Gaussian states with Gaussian operations}

The availability of distillation protocols will be of crucial
importance in the infinite-dimensional setting when we wish to
implement long-distance quantum communication based on continuous
variables. In fact, such distillation protocols must be performed
at the beginning of any procedure that relies on the availability
of highly entangled approximately pure shared entangled states.
Noise due to unwanted coupling to an environment is never entirely
avoidable, and protocols have to be devised that effectively
reverse this process. In the finite dimensional setting,
distillation protocols are known, and using the polarization
degrees of freedom of light, they can be implemented optically
(although they require photon counters in an iterative procedure
that distinguish with great efficiency different numbers of
photons) \cite{Zeilinger}.

\begin{figure}[ht]
\centerline{
       \epsfxsize=\columnwidth
       \epsfbox{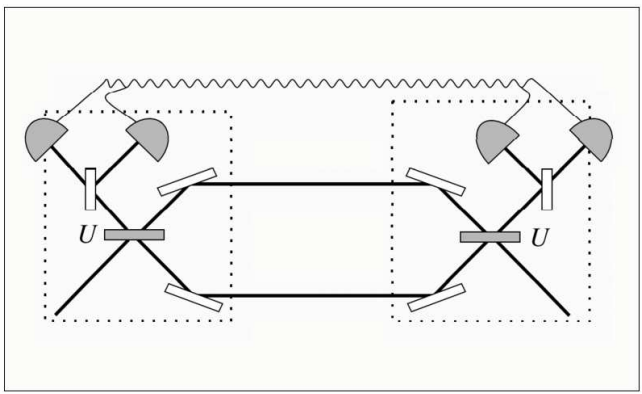}
       }
\smallskip

\caption{The class of considered feasible Gaussian distillation
protocols.}\label{DGauss}
\end{figure}

 At the time it seemed fairly  natural to expect that such
distillation schemes can also be constructed in the case of
Gaussian states and Gaussian operations. Let us hence for a moment
assume that pure Gaussian two-mode squeezed states have been
transmitted through lossy optical systems such as optical fibers.
The resulting states are still Gaussian, provided that  the noisy
channel was a Gaussian channel, as is a good assumption in case of
optical fibers. In order to distill the entanglement in an
iterative procedure, one would take two pairs of identically
prepared systems $\rho\otimes \rho$ and would feed them into one
step of the procedure: the output could then be taken as the input
of the next step. The states are identical if they underwent the
same losses. The corresponding modes will from now on be denoted
as $A1$, $A2$, $B1$, and $B2$. A feasible iterative distillation
protocol preserving the Gaussian character of an input state would
look as follows (see Fig.\ \ref{DGauss}):
\begin{itemize}
\item[(i)] Application of a local Gaussian unitary operation, i.e., a map of the form
\begin{equation}
    \rho\otimes \rho\longmapsto (U_A\otimes U_B)(\rho\odot \rho)
    (U_A\otimes U_B)^\dagger.
\end{equation}
 Note that the $\odot$ denotes the tensor product between
different copies, while the $\otimes$ denotes the tensor product
between the two parties. As has been discussed
before, this set includes the passive transformations embodying
those operations that can be implemented in optical systems using
beam splitters and phase shifts. It however also includes
squeezing operations. For the two-mode case, the set of local
unitary transformation is a 20-dimensional manifold. Note that it
is not required that both parties implement the same operation.

\item[(ii)] A homodyne measurement on the modes $A2$ and $B2$.

\item[(iii)]Finally,
the parties communicate classically about the outcome of the
measurement, and perform postprocessing of the states of modes
$A1$ and $B1$ with unitary Gaussian operations.
\end{itemize}

It was expected that Gaussian operations would be sufficient to
provide for such an entanglement purification protocol. It was
however proven in Ref.\ \cite{Eisert SP 02b} that this is not the
case. Later this proof was extended to demonstrate that
entanglement distillation of Gaussian state with Gaussian
operations is not possible in general, allowing in particular for
a $2n$-mode input \cite{Giedke C 02}. This is after all quite an
astonishing result: no matter how the Gaussian local operation
with classical operation is chosen, the degree of entanglement can
not be increased. Whatever operation is implemented, it will
result in a loss rather than in an increase of entanglement. The
optimal procedure is simple: not to even try to distill with
Gaussian operations alone. This remarkable result also implies
that quantum error correction is not feasible if one is restricted
to Gaussian states and Gaussian operations. Hence, it becomes
clear that is necessary to add some non-Gaussian resource to the
repertoire to ensure that interesting quantum information
processing tasks can be carried out in the presence of noise.

The statement has been based on the quantification on the degree
of entanglement in terms of the logarithmic negativity that has
been mentioned before, which is defined as
\begin{equation}
    E_{N}(\rho)=
    \log_2 \|\rho^{T_{A}}\|_1
\end{equation}
for a state $\rho$, where $\|.\|_1$ again denotes the trace norm,
and $\rho^{T_{A}}$ is the partial transpose of $\rho$. For
Gaussian states it has an interpretation in terms of a certain
asymptotic entanglement cost \cite{Audenaert PE 03}. In the
meantime, it should be noted, the entanglement of formation is
also available for the case of symmetric two-mode Gaussian states,
i.e., states for which the two reductions of the covariance matrix
are identical up to local symplectic transformations \cite{Wolf
GKWC 03}.
For pure (and for symmetric mixed) Gaussian states the logarithmic
negativity is related to the degree of squeezing in a monotone way
(see, e.g., Ref.\ \cite{Wolf EP 03}).  These considerations, hence,
imply again that with Gaussian operations alone two identically
prepared two-mode squeezed states cannot be transformed into a
single two-mode squeezed state with a higher degree of squeezing,
as has already been pointed out in Eq.\ (\ref{MainCrit}).

\subsection{Minimal non-Gaussian extensions for distillation}

This difficulty sketched in this section -- the impossibility of
distilling Gaussian states with Gaussian operations -- can however
be overcome. It has been known for quite some time that the
unlimited availability of arbitrary non-Gaussian operations
essentially requiring quantum computation allow for example for
entanglement distillation of Gaussian states \cite{Parker BP
00,Giedke DCZ 01} to finite-dimensional singlets. This is an
important result from the perspective of the theory of
entanglement. Yet, not only from a practical point of view it
would be crucially important to find solutions that make merely
use of minimal extensions of the set of operations into the
non-Gaussian regime, including only those additional operations
that are to a high extent feasible in quantum optical settings.

In a nutshell, it turns out that it {\em is} possible  indeed to
distill continuous-variable entanglement yielding highly entangled
approximately Gaussian states, and this can be done in optical
systems using passive optical elements and photon detectors only.
Such procedures have been introduced and investigated in detail in
Refs.\ \cite{Browne ESP 03,Eisert BSP 03}. The key idea is to
leave the Gaussian setting in a first non-Gaussian step which
involves a single measurement only (corresponding to the positive
outcome of a photon detector), and then apply Gaussian operations
only.
The philosophy of the protocol is sketched in the following:
Some source provides a two-mode squeezed state which is
transmitted through fibers whose imperfections turn the states
into symmetric noisy but still Gaussian states. At that stage a
single non-Gaussian operation is carried out which takes the
states out of the Gaussian state space and may perhaps also
increase entanglement. Subsequent to that only Gaussian operations
are applied in an iterative scheme which serve two aims. Firstly,
they further increase the amount of entanglement and secondly,
they make the states progressively more Gaussian. Asymptotically
the states may then have a higher degree of entanglement than the
original supply and at the same time approach Gaussian states as
closely as possible.
\begin{figure}[ht]
\centerline{
       \epsfxsize=\columnwidth
      \epsfbox{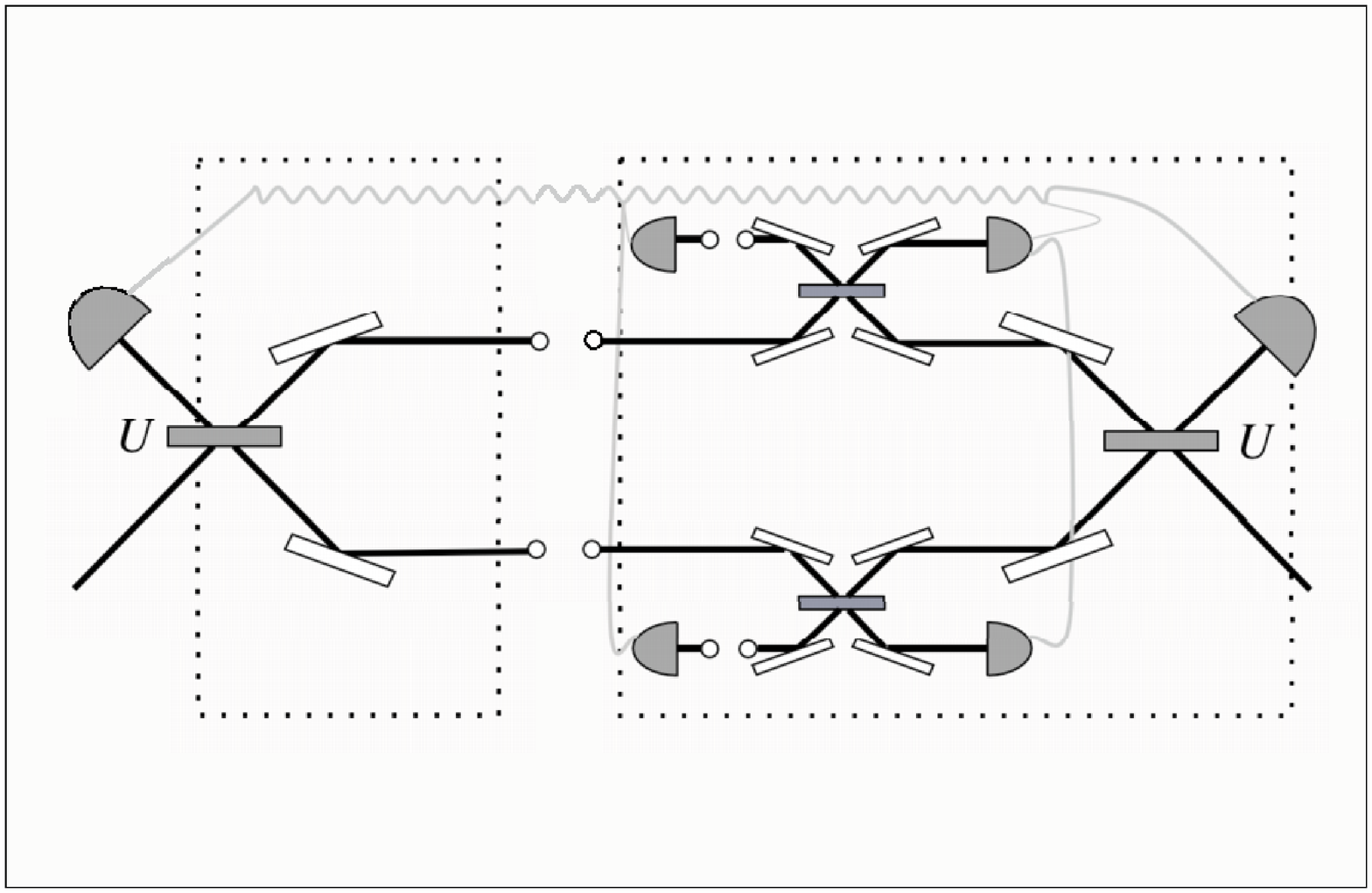}}
\smallskip
\caption{\label{Gauss} A possible scheme for entanglement
distillation of Gaussian states using minimal non-Gaussian
resources. Gaussian entangled states are corrupted during
transmission but retain their Gaussian character. A single
non-Gaussian operation is applied taking the states out of the
Gaussian state space. Subsequently only Gaussian operations are
applied which both distill entanglement and drive the states
closer to the set of Gaussian states. }
\end{figure}

Surprisingly indeed, such a scheme exists using as the only
non-Gaussian resource a photo-detector that can distinguish
between the presence and absence of photons but is not able to
count the precise number of photons (yes-no detector) \cite{Browne
ESP 03,Eisert BSP 03}. The first non-Gaussian step takes two
identical copies of two-mode squeezed states, and measurements are
performed on two of the four modes. The outcome is accepted as
successful in case that both detectors click (i.e., register at
least one photon), corresponding to the Kraus operators $E_2
=\id-|0\rangle\langle0|$. The reflectivity and transmittivities of
the beam splitter $V$ are not $50:50$, but have to be tuned
appropriately. The following iterative Gaussian procedure takes
two pairs $\rho\odot \rho$ of non-Gaussian states $\rho$ as input
which are then mixed at a $50:50$ beam splitter,
where the transformation induced by the beam splitters are given
by unitaries
\begin{equation}
    U= T^{n_{1}} e^{-R^{\ast}a_{2}^{\dagger} a_{1} }
    e^{R a_{2} a_{1}^{\dagger}}
    T^{-n_{2}},
\end{equation}
with $T=R=1/\sqrt{2}$. Two of the output modes are then fed into a
photon detector, each associated with Kraus operators
\begin{equation}
    E_1 = |0\rangle\langle 0|,\,\,
    E_2 = \id -|0\rangle\langle 0|,
\end{equation}
where $|0\rangle$ denotes the state vector associated with the
vacuum state. The state is kept in case of the vacuum outcome of
both local detectors. The unnormalized final state after one step
is hence given by
\begin{eqnarray}
    \rho'= \langle 0,0| (U\otimes U) (\rho\odot \rho)(U\otimes U)^{\dagger}
    |0,0\rangle.
\end{eqnarray}
The resulting two-mode states then form the basis of the next
step. It is an iterative protocol, and it is event-ready, in the
sense that one has a classical signal at hand which indicates
whether the procedure was successful or not. No further
post-processing has to be performed. Generic weak convergence to
two-mode Gaussians can indeed be proven, and it can be
demonstrated that the procedure often leads to highly entangled
and squeezed states. Remarkably indeed, the scheme turns out to be
fairly robust with respect to detector imperfections \cite{Eisert
BSP 03}. It is beyond the scope of this paper to present details
of this procedure, however, and the reader is referred to the
literature here.


\subsection{Generation of entanglement with passive transformations}

The
previous considerations were concerned with the local manipulation
of Gaussian states. This is a meaningful approach if one has a
distributed system, where the necessity of local operations is
simply dictated by the set-up. But how does one prepare the
entangled states in the first place? Experimentally, there are
several ways to do it, and again, it would be beyond the scope of
this paper to present these possibilities (see, e.g., Ref.\
\cite{Korolkova SGLML 02}). Instead, to exemplify the formalism
developed before, we concentrate on the optimal creation of
entanglement of Gaussian states using passive transformations only
\cite{Wolf EP 03,Korolkova SGLML 02,Kim SBK 02,Enk 03}. So in this
subsection, the operations are allowed to be non-local with
respect to the modes that have to be entangled. Passive operations
are, again in optical systems singled out as they correspond to
those operations that can be realized with beam splitters and
phase shifts. If simply the vacuum is fed into the additional
input modes, then one even faces no practical problems of mode
matching in optical systems.
To start with, if one intends to entangle $n$ Gaussian input modes
with passive transformations, according to
\begin{equation}
    \gamma\longmapsto \gamma'= S\gamma S^T ,\,\,S\in K(2n),
\end{equation}
the input states have to be  squeezed. This is immediately
obvious: any state that is entangled must be squeezed, since
otherwise, the covariance matrix $\gamma'$ of the $n$-mode output
would satisfy
\begin{equation}
\gamma' \geq {\id}_{2n}\oplus \id_{2n},
\end{equation}
which in turn according to (\ref{generalcrit}) implies
separability. But what is the optimal operation, and, first and
foremost, what is the optimal degree of entanglement that can be
achieved in any two-mode output?

This question can actually be completely solved, when the degree
of entanglement is quantified in terms of the negativity
\cite{Wolf EP 03}. Let $\gamma$ be the covariance matrix of the
$n$-mode input state, Gaussian but not necessarily pure, and
$\gamma'= S\gamma S^T$ be the resulting covariance matrix after
application of the passive transformations. Then, the maximum
amount of entanglement obtained for any two-mode sub-system
is given by
\begin{equation}
    E_N(\rho) =\max\left(0,\log (\lambda_1\lambda_2)/2\right),
\end{equation}
 where $\lambda_1$ and $\lambda_2$ are the two smallest
eigenvalues of $\gamma$. This general solution of the problem
establishes a generically valid link between the squeezing of the
initial state to the degree of entanglement that can be
potentially unlocked with passive transformations.  The proof of
this statement as found in Ref.\ \cite{Wolf EP 03} makes extensive
use of the isomorphism between the groups $K(n)$ and $U(n)$. The
optimal entangling transformation can moreover be constructively
found, and in the practically important case of two input modes, a
formula can be given for the appropriate choices of
transmittivities, reflectivities, and phases of the optical
elements.

\section{Conclusions}
In this article we have briefly reviewed some elements of the
theory of entanglement both in finite and infinite dimensional
systems. We have formulated the basic questions that any theory of
entanglement aims to answer when we consider entanglement as a
resource for quantum information processing protocols. These key
questions are those of characterization, (local) manipulation, and
quantification of the entanglement resource. We then pointed out
that entanglement theory in finite dimensional systems such as
qubits generally requires the availability of arbitrary local
operations. Statements concerning the feasibility or efficiency of
a particular local state transformation may require the use of
operations that are extraordinarily difficult to realize in
practice. This motivated the development of a theory of
entanglement under experimentally available operations and a
particularly relevant example of such a theory can be found in
infinite dimensional systems that are equivalent to harmonic
oscillators. Examples of such systems are the photonic degree of
freedom of a light mode, nano-mechanical oscillators and,
approximately, cold atomic gases. The restricted set of
experimentally available operations that is being considered in
this theory is that of Gaussian operations which allow for the
creation and manipulation of Gaussian states. This restricted set
of operations is of particular importance in quantum optics where
it can be shown that Gaussian operations are in one-to-one
correspondence to the set of operations that can be implemented by
simple optical tools such as phase plates, beam-splitters, and
squeezer together with the addition of vacuum modes and homodyne
detection. In the remainder of the article we then outlined the
recent development of this theory and stated some of its key
results. In particular it became clear that processes such as
quantum error correction, entanglement distillation and efficient
quantum computation are not possible under this restricted set of
operations but require some resources that lie outside the
Gaussian regime. This suggested to search for protocols that
employ non-Gaussian resources in a minimal way. Indeed, such
extensions are possible in the sense that the expensive
non-Gaussian operations are used only in an initial step of the
protocol and are then followed by purely Gaussian operations.

Finally it should be mentioned that the recent development of the
theory of entanglement for Gaussian states has also provided us
with many novel and useful techniques for the analytical study of
entanglement in harmonic systems. This permits both, to revisit
old questions such as for example concerning the connection
between entanglement and the dynamical appearance of classical
properties  in quantum Brownian motion \cite{Eisert P 02b}, but
also newly arising  questions concerning the study of entanglement
properties of interacting quantum systems both in the static case
\cite{Audenaert EPW 02} and in their dynamical behaviour
\cite{Eisert PB 03,Plenio BEH 03}. In these problems, the theory
of Gaussian entangled states often allows for the exact analytical
solution of many questions and therefore provides an ideal
playground for the exploration of quantum entanglement.

\section*{Acknowledgements} Discussions and collaborations on
the subject presented here with K.\ Audenaert, D.\ Browne, K.\
Banaszek, S.\ Bose, J.I.\ Cirac, G.\ Giedke, Ch.\ Silberhorn, M.\
Lewenstein, M.M.\ Wolf, S.\ Scheel, I.A.\ Walmsley, and R.F.\
Werner are gratefully acknowledged, as well as constructive
remarks on the manuscript by J.\ Anders. The research described
here has been partly supported by the A.-v.-Humboldt Foundation,
the ESF program "Quantum Information Theory and Quantum
Computation", EPSRC, the European Union via projects EQUIP,
QUPRODIS, and QUIPROCONE and a Royal Society Leverhulme Trust
Senior Research Fellowship.


\begin{thebibliography}{99}
\bibitem{Commentonchoiceofnorm}
Note that in the infinite setting different norms do not
necessarily imply the same topologies. However, the choice of
trace norm is quite natural as two states that are close in the
trace norm will yield close expectation values for bounded
observables or, in other words, the more difficult it is to
operationally distinguish the states with generalized measurements
in the non-asymptotic regime.
\end{thebibliography}
\end{document}